\documentclass[12pt,preprint]{aastex}
\usepackage{emulateapj5}
\usepackage{graphicx}
\usepackage{psfig}
\usepackage{natbib}
\slugcomment{draft version}

\begin{document}


\title{X-ray spectroscopy of PSR B1951+32 and its pulsar wind nebula}


\author{X.H. Li\altaffilmark{1}, F.J. Lu\altaffilmark{1},
  T.P. Li\altaffilmark{1,2,3}} 


\altaffiltext{1}{Laboratory of Particle Astrophysics, 
Institute of High Energy Physics, CAS, Beijing 100049, P.R. China;
lixh@ihep.ac.cn; lufj@ihep.ac.cn}
\altaffiltext{2}{Department of Physics \& Center for Astrophysics, Tsinghua
University, Beijing; litp@mail.tsinghua.edu.cn}
\altaffiltext{3}{Department of Engineering Physics \& Center for
Astrophysics, Tsinghua university}

\begin{abstract}
We present spatially resolved X-ray spectroscopy of PSR B1951+32 and
its pulsar wind nebula (PWN) in supernova remnant (SNR)  CTB 80 using a
{\sl Chandra} observation. The {\sl Chandra} X-ray map reveals 
clearly various 
components of a ram-pressure confined PWN embedded in the SNR ejecta: a
point source representing the pulsar, X-ray emission from the bow shock,  
a luminous X-ray tail, a 30$\arcsec$ diameter plateau whose 
northwestern part is absent, and the outside more diffuse X-ray
emission. The plateau is just surrounded by the radio, [O III], 
[S II], and [N II] shells, and the outside diffuse emission
is mostly within the H${\alpha}$ shells.  While the spectra of 
all the features are well fitted with power law models,  a power 
law plus blackbody model can fit the spectrum of the pulsar significantly
 better than using a power law model alone.  Generally 
the spectra of these components obey the trend of steepening from 
the inside to the outside. However, the edge of the plateau probably has
a harder spectrum than that of the central region of the plateau. 
The cause of the apparent hard  
spectrum of the plateau edge is unclear, and we speculate that it might
be due to a shock between the PWN and the SNR ejecta.      
The possible blackbody radiation component from the pulsar 
has a temperature 
of 0.13$\pm0.02$ keV and an equivalent  emitting radius of  
 2.2$^{+1.4}_{-0.8}$ (d/2 kpc) km, and is thus probably from the hot spots on 
the pulsar. We also show in this paper that the blackbody temperature
of the entire surface of PSR B1951+32 is much lower than those 
predicted by the standard neutron star cooling
models.
 
\end{abstract}


\section{Introduction}
It is generally accepted that only a small fraction ($\le$ 10\%) of the
 spin-down energy of a young pulsar is converted into observable
 pulsed emission, and most of the energy leaves the pulsar
 in the form of a  highly relativistic electron/positron
 wind. This relativistic wind eventually interacts (through a
 termination shock)  with the external medium,
 emits synchrotron radiation, and produces a pulsar wind nebula (PWN) 
\citep[e.g.,][]{Rees1974, Kennel1984}. 
The pulsar wind particles are 
usually confined by the supernova ejecta or 
interstellar medium (ISM) \citep[e.g.,][]{Reynolds1984}, and the 
morphology of the PWN relies strongly on how the wind particles 
are confined.  Some PWNe (e.g., IC 433, W44, N157B, B1957+20, B1757-24, 
Duck nebula, Mouse, Geminga) have 
cometary morphologies, which are suggested to be due to the supersonic motion
of their respective pulsars in the ISM or supernova remnants (SNRs) 
\citep{Wang1993, Olbert2001, Wang2001, Kaspi2001, Chatterjee2002, Petre2002,
 Caraveo2003,  Lu2003, Stappers2003, Gaensler2004, Gvaramadze2004}. 
In such a system, there is a bow shock running ahead of the pulsar,
most of the wind particles are confined to the direction opposite 
to the pulsar proper motion, and the pulsar is well offset from
the center or even at the apex of the PWN  \citep{Wang1993, Wilkin1996, 
Wang1998, Bucciantini2002,  Caraveo2003, Swaluw2003, Gaensler2004, 
Gvaramadze2004, Swaluw2004}. 
Detailed spatially resolved X-ray spectroscopy of
a PWN should be important for examining and refining those 
models, because the X-ray emitting particles are young
and have a short lifetime, and thus the spatial variations of the
X-ray spectra reflect very well the particle track and  energy 
evolution in the PWN. 

Radio and  optical  observations suggest that the PWN energized by 
PSR B1951+32 in SNR CTB 80 (G69.0+2.7) is an ideal
laboratory to study the interaction between an SNR and the embedded PWN. 
CTB 80 has a peculiar  radio morphology composed of three ridges and 
a $\sim 30\arcsec$ core in the
southern central portion of the extended component
\citep[e.g.,][]{Koo1993, Mav2001, Cas2003}.  
The 39.5 ms pulsar PSR B1951+32, located at
the south-western edge of the core, moves towards the 
south-west ($\sim252\degr$, north through east) with a 
transverse velocity  of 240$\pm$40 km s$^{-1}$ for a distance 
 of 2 kpc \citep{Mig2002}. Radio observation revealed
 a ``$U$'' like loop and a bright bow shock feature at the
southwest of the loop, indicating 
strong interaction of the wind of the fast moving 
pulsar with the environment \citep{Strom1987, Mig2002}. 
The optical structure of the PWN of PSR B1951+32
was delineated by forbidden-line ([O III], [S II], and [N II])
emission as shell-like \citep{HK1988, HK1989}. 
These optical features suggest that they arise behind shocks which are
being driven into a magnetized thermal plasma by the confined relativistic
wind from PSR B1951+32 \citep{HK1989}. 

The optical and radio properties make the PWN of PSR B1951+32 also a hot 
target for various X-ray telescopes. 
The {\sl EINSTEIN} observations show a central filled X-ray source
with nonthermal X-ray spectrum (photon index $\alpha$=3.8$^{+0.2}_{-0.3}$), 
and the morphology of the source was suggested as a result of the 
relativistic jets energized by the central
pulsar \citep{Becker1982, Wang1984}.  The 
{\sl EXOSAT} observation confirmed the nonthermal nature of the 
X-ray spectrum but inferred a smaller photon index of 1.9$\pm0.5$
(Angelini et al. 1988).  
Safi-Harb et al. (1995) studied CTB 80 and
PSR B1951+32 with the Position Sensitive Proportional Counter (PSPC) and
the High Resolution Imager (HRI) onboard {\sl ROSAT}.  They found
 a bright compact core of $\sim$ 1$\arcmin$
radius surrounding the pulsar and a diffuse nebula
extending $\sim5\arcmin$ eastward of the pulsar, and the spectra of 
these two features are both nonthermal.  
These observational properties are well consistent with those of a
PWN. However, due to the limited spatial and spectral resolution 
of the previous X-ray telescopes, the detailed 
morphological and spectral structures of the X-ray emission remains
not well resolved.

The superb spatial resolution and moderate spectral resolution of
the {\sl Chandra X-ray Observatory} (hereafter {\sl Chandra}) 
permit a detailed morphological study
and spatially resolved X-ray spectroscopy of PSR B1951+32 and its PWN.
{\sl Chandra} can isolate the pulsar from the surrounding 
nebula and we can then study the spectrum of the pulsar proper, which
has not been done so far yet. The lifetime of the synchrotron 
X-ray emitting particles is 
short, and therefore the spectral variation across the nebula presents 
important clues to the particle acceleration and the 
energy losing processes.    
Recently, Moon et al. (2004) studied the high resolution 
X-ray (with {\sl Chandra}), H${\alpha}$ (with {\sl Hubble}), and IR 
(with {\sl 5-m Palomar Hale telescope}) structures of the region around 
PSR B1951+32 and revealed a cometary PWN which appears
to be confined by a bow shock produced by the high-velocity motion of the 
pulsar. In this paper, we give more detailed analyses of the {\sl Chandra}
data.  We first introduce the data reduction in
 $\S$ 2, then present our analyses and results in $\S$ 3, discuss the
 structure of PWN in $\S$ 4, and conclude our work in $\S$ 5.  All
through the paper, the errors are at the 90$\%$ significance level.  

\section{Observation and Data Reduction}
{\sl Chandra} observed the PWN of PSR B1951+32 by the Advanced CCD Imaging
Spectrometer (ACIS)  on July 19th,
2001 with an exposure time of 74 ks. 
The target was positioned at the aim-point on the back illuminated
ACIS-S3 in the ``VFAINT'' mode and at a working temperature of
$-$120$\degr$C. 
 ACIS is sensitive to X-rays in 0.2-10 keV with an
energy  resolution of $\Delta E/E \sim 0.1$ at 1 keV, and 
the full width at half maximum (FWHM) 
of the 
point spread function (PSF) is 0$\farcs5$. The frame read out time
for this observation is 0.74 s, since only a small portion of the 
CCD chip was illuminated.

We calibrated the data using {\it CIAO} (version 3.1) and its CALDB
(version 2.27). 
We reprocessed the Level 1 data for correction of the charge transfer
inefficiency (CTI) effects, cleaned the background, and removed
the afterglow. Time intervals with
anomalous background
rates associated with  particle flare events were further rejected for
the Level 2 data, and the final net exposure time is 71 ks.
The spectra were fitted with {\it XSPEC}.

\section{Analysis and Results}
\subsection{Spatial structure}
Figure 1 shows an ACIS image of the PWN of PSR B1951+32.
 This image reveals several major 
components of the PWN: a point source at  
RA (J2000) = $\rm 19^{h}52^{m}58\fs20$, DEC (J2000) =
$32\arcdeg52\arcmin40\arcsec.7$,  
a bright elongation lies in the northeast of the point source,
a 30$\arcsec$ diameter plateau with absence in its northwest, 
and more diffuse emission  
in between and beyond these features. 
The position of the point source was obtained by the {\it celldetect} tool in 
{\it CIAO} and has an uncertainty  $\sim$0$\farcs$2, which is well 
consistent with the radio position of PSR B1951+32 \citep{Mig2002}.  
The X-ray point source thus represents the X-ray emission from this pulsar.
The X-ray plateau is just within the radio and optical 
shells \citep{HK1989}, and therefore
corresponds to the main body of the PWN.  The overall structure of the
 X-ray nebula is similar to the radio structure, as shown in figure 2,
 except that the radio nebula is limb-brightened and that the bright X-ray
 elongation in the northeast is absent in the radio map 
 \citep{Strom1987,Mig2002}.   

In order to show the diffuse X-ray emission near the pulsar more clearly,
we plot in Figure 3 the X-ray counts (per 0$\farcs492\times0\farcs492$
pixel) profile (solid line) along the  
pulsar proper motion. In this profile the contribution from the pulsar has been
removed by subtracting the convolution of a delta function with 
the telescope PSF. The PSF is energy-weighted and was simulated with 
{\it  ChaRT}{\footnote{http://cxc.harvard.edu/chart/threads/index.html}}. The  
dashed line in 
Figure 3 represents the 1.5 GHz radio profile from \cite{Mig2002}. 
It is clear
that there is high brightness diffuse emission within $\sim$2$\arcsec$ radius
from the pulsar. There is also significant X-ray emission in the 
radio bow shock region, although the overall trend of the diffuse 
emission is declining.  

\subsection{Spectra}
According to our analyses of the morphology of the nebula,
we divided it into a few regions (see Figure 4), from which 
we extracted the spectra of various components. The background used in
the spectral  
analysis of the diffuse features was extracted from two boxes 
where there are no prominent emission, while the background 
of the pulsar spectrum was extracted from an annulus centered
on the pulsar, with inner and outer radii of 1$\farcs5$ and 
3$\arcsec$ respectively. We fitted the
 spectra of the diffuse features jointly with a
power law (PL) model by forcing all the spectra to share the same absorbing 
column density ($N_{\rm H}$). As shown in Figure 5, such a model fits 
the spectra very well.
The resulted $N_{\rm H}$ is 3.0$\pm0.1\times10^{21}$ cm$^{-2}$,
and the other parameters are listed in Table 1.  As seen in other PWNe, these 
spectra have a trend of softening with distance from the pulsar due to
the fast energy-losing of the high energy electrons
\citep[e.g.,][]{Slane2002,Lu2002,Kaspi2003}.

The pile up fraction of the pulsar emission was estimated as 6$\%$ by using
{\it sherpa}, and thus the pile up effect was neglected in our analysis
of the pulsar spectrum.
We first fitted the pulsar spectrum with a PL model by fixing
the column density to 3.0$\times10^{21}$ cm$^{-2}$ as derived
above. This gives a photon index of 1.74$\pm0.03$ and $\chi^2$ 
of 223 with 208 degrees of freedom (dof). We then fitted the spectrum with 
 a power law + blackbody (PL+BB) model, and this yields a photon
 index of  1.63$^{+0.03}_{-0.05}$,  
BB temperature of 0.13$\pm0.02$ keV, and a $\chi^2$
of 190 with 206 dof. If we only check the data in 0.5-1.5 keV
(to which energy region the BB component contributes), the $\chi^2$ 
and dof of the spectral fitting with and without the BB component
are 61, 60 and 84, 62, respectively. $ftest$ shows that 
the substitution of 
the PL model by the PL+BB model is 
necessary at a significance level of $>$ 99.99\%. In comparison, Figure
6 gives the 0.5-1.5 keV spectrum of the pulsar fitted  
with a PL model and a PL+BB model. 
The fitted
BB flux corresponds to an equivalent emitting radius of
2.2$^{+1.4}_{-0.8}$ km assuming a distance of 2.0 kpc to the pulsar. 
 
In order to see whether there is any spectral variation in the plateau
and in the very outside region surrounding the plateau, 
we extracted spectra from the 5 quasi-annulus
regions defined in Figure 7 and fitted them with a PL model 
using the background  
identical to that used above. Table 2 lists the spectral
fitting results, which show that the edge of the plateau 
(rings 1, 2, and 3 in Table 2, and so the radio shell region) 
has apparently a harder spectrum  
than the neighboring inner (the center region) and outer regions (the
ring 4). 

A monte-carlo simulation has been used to estimate how significant the 
spectrum of the plateau edge is flatter than 
that of the center region. We generated
100,000 pairs of random numbers. In each pair, the first is a random
number from a gaussian distribution with mean and standard deviation
of 1.76 and 0.04, while the second number is from a gaussian distribution
with mean and standard deviation of 1.64 and 0.015. 
The standard deviations are smaller than the errors listed in Table 2 since 
those in Table 2 are at 90\% significance level. We found that 
only 235 out of the 100,000 pairs have their first numbers smaller than the
corresponding second numbers. Therefore, 
the overall spectrum of rings
1 to 3 is  flatter than the spectrum of the center region at a significance of 
 99.7\%. 

We have also fitted the spectrum of the entire nebula (excluding the pulsar)
with a power law model. This gives an $N_{\rm H}$ of 
3.0$\pm0.1\times10^{21}$ cm$^{-2}$, a photon index of 1.73$\pm0.03$, an 
unabsorbed 0.2-10 keV 
energy flux of 9.6$\times10^{-12}$ ergs cm$^{-2}$ s$^{-1}$, and  
$\chi^2$ of 296 for 312 dof. 
 
\section{Discussion}
\subsection{The magnetic field strength in the nebula}

The radio and X-ray spectra of the PWN can be used to estimate the 
magnetic field strength in the nebula. 
The radio emission of this PWN 
has a flat spectrum of $\alpha=0$ ($F_{\nu}\sim\nu^{-\alpha}$) and
flux of $\sim$500 mJy \citep{Angerhofer1981}. From $\S$ 3.2 we 
inferred these two values for the X-ray emission as 0.73 and 1 $\mu$Jy,
respectively. Therefore 
the spectral break frequency $\nu_{br}$ between the radio and X-ray band is 
around 10$^{10}$ Hz. Assuming that this spectral break arises from 
synchrotron losses, the magnetic field in the nebula is then determined
by the age of this nebula. 
 The proper motion of the pulsar indicates that the pulsar can cross the 
X-ray plateau in about 1200 yr.
Nonetheless, this seems unlikely the age of the PWN of PSR B1951+32.
Most of the electrons injected by the pulsar in the history are probably still
confined in the bubble by the highly ordered magnetic field \citep{HK1988}, as indicated by the much higher radio
brightness of this  region (than the neighboring regions) and the
nice radio shells \citep{Mig2002}. Therefore, we use  the
pulsar dynamic age (64 kyr, Migliazzo et al. 2002) to represent 
the age of the PWN. 
Using the formula reproduced by \cite{Frail1996} from Pacholczyk
(1970), we derived that the magnetic field strength in
the nebula is $\sim 300 \mu $G. 
This estimated magnetic field strength is comparable with the
values given by \cite{Angerhofer1981}, \cite{HK1989}, and \cite{Moon2004}, 
but much higher than those by \cite{Safi_harb1995} and \cite{Cas2003}.

The discrepancy between our result and that (5.2 $\mu$ G) of \cite{Cas2003} is
due to that they used a much higher $\nu_{br}$ of
2.4$\times10^{16}$ Hz and a much lower PWN age of 18,200 yr. We found that 
both the high $\nu_{br}$ and the low PWN age are problematic. 
Their reason of choosing $\nu_{br}$ = 2.4$\times10^{16}$ Hz is
that $\nu_{br}$ is not much lower than the X-ray energies given
the similar X-ray and radio sizes of the PWN. However, as we discussed
above, the PWN has the similar X-ray and radio sizes most probably
because the pulsar wind particles are well confined rather than  
diffuse freely. 18,200 yr is the time for the pulsar to move across
the $\sim$10$\arcmin$ radio nebula. But, again, since most of the 
wind particles ejected by the pulsar in the history are likely still
in the bubble. To use the pulsar dynamic age  as the PWN age is more
reasonale.  

We now discuss  whether the spectral break
between the radio and X-ray bands is mainly due to synchrotron cooling.
 \cite{Kaspi2001} suggested that 
more than one electron populations are  
being injected into the PWN of PSR B1757-24. The injected wind particles
of PSR B1951+32 may have similar energy spectrum. However, 
synchrotron cooling does play an important role in the X-ray to 
radio spectrum in the PWN of PSR B1951+32. The particle 
energy spectrum of the bright X-ray tail (see Figures 1 and 3) may 
represent the injected particle energy spectrum well, 
because these particles are young. The non-detection of radio emission
in this region suggests that the current overall particle energy spectrum 
in the PWN is significantly steeper than the injected one, due to 
synchrotron cooling. Therefore, although not very certain, 
the magnetic field strength we obtained is reasonable. 
  
\subsection{The PWN geometry}

The PWN of PSR B1951+32 represents the pulsar wind material confined by
the ram pressure of the ambient medium,
a typical example of the model suggested  by many 
authors \citep[e.g.,][]{Wilkin1996, Wang1998, Bucciantini2002, Swaluw2003, 
Gaensler2004, Swaluw2004}. It has the following 
regions:
\begin{itemize}
\item{\bf The pulsar wind cavity and termination shock:} 
In the immediate region surrounding the pulsar the relativistic 
wind flows freely outwards with very little emission,  and  
this region is shown as a cavity. The size of 
the cavity is determined 
by the radius of the termination shock 
over where the pulsar wind 
ram pressure ($\frac{\dot{E}}{4\pi r^2 c}$) balances the 
ram pressure of the ISM due to the
motion of the pulsar ($\rho_{\rm a} v_{\rm psr}^2$),
where $\rho_{\rm a}$ is the ambient medium density, $v_{\rm psr}$ and $\dot{E}$ the velocity and
spin-down energy of the pulsar, $r$ the radius of the termination shock, and $c$ the speed
of light. For PSR B1951+32, 
$\rho_{\rm a} \sim 50\times 1.67 \times10^{-24}$ g cm$^{-3} = 8.4\times10^{-24}$ g cm$^{-3}$\citep{HK1989},
$v_{\rm psr}\sim$ 240 km s$^{-1}$, $\dot{E}=3.7\times10^{36}$ erg s$^{-1}$, and 
thus $r\sim 1.4\times10^{\rm 16}$ cm, corresponding to 0.5$\arcsec$ at the distance (2.0 kpc)
to the pulsar. Because PSR B1951+32 moves
supersonically in the SNR, as the simulation shows
\citep[e.g.,][]{Gaensler2004}, 
this region is more likely elongated opposite to the pulsar proper motion
rather than spheric. 

In the X-ray counts profile of the nebula along the pulsar proper motion
(Figure 3) we see strong diffuse emission in the  nearby region 
of the pulsar. The FWHM of this emission is about
3$\arcsec$, and the emission following the pulsar  
extends a little further out ($\sim3\arcsec$ compared to $\sim 2\arcsec$) than 
that ahead of the pulsar.  We propose that  this high brightness region 
represents most probably 
the emission from the termination shock, giving their roughly 
similar angular sizes. 
The bigger extension opposite to the pulsar proper motion is
consistent with the distorted termination shock geometry 
because of the supersonic motion of the pulsar
\citep{Gaensler2004}. The pulsar wind cavity in this nebula is
unresolved, giving the contamination from the high brightness emission
from the pulsar and the termination shock.

\item{\bf The particle tunnel:} 
The X-ray elongation (tail) about 4$\arcsec$ to 10$\arcsec$ northeastern from
the pulsar represents most probably the particle tunnel suggested by
\cite{Wang1998} and \cite{Gaensler2004} (Region B2 in Figure 9 of
Gaensler et al's paper). Reasons to identify this emission as from the 
particle tunnel rather than from the termination shock are the tail like
morphology and its much lower brightness than that of the region
neighboring the pulsar.  
The bright radio and X-ray features at
 RA (J2000) = $\rm 19^{h}52^{m}59\fs6$, 
DEC (J2000) = $\rm 32\degr52\arcmin44\arcsec$ 
\citep[Figure2;][]{Strom1987,Mig2002} may represent the termination
site of the particle tunnel \citep{Wang1998}.  Since this bright radio
feature does not align with the pulsar's motion, it is possible
that the particle tunnel has been bent due to 
interaction with a gradient in the ambient density and/or magnetic
field, as also seen in the Crab nebula 
and the PWN of PSR B1509-58 \citep{Weisskopf2000, Gaensler2002}.

\item{\bf The interface between the PWN and SNR:}
In $\S$ 3.1 we have mentioned that bright X-ray plateau are well surrounded 
by the radio and [O III], [S II] and [N II] shells. This implies that the 
main body of the PWN is well confined by the SNR ejecta. The radio map
is much more shell-like than the X-ray map, suggesting that the old wind 
particles have accumulated at the interface between the PWN and SNR, 
since the radio emitting particles have a much longer lifetime than 
 the X-ray emitting particles. As suggested by \cite{HK1988}, 
there may exist highly ordered magnetic field in the shell, which might 
prevent the old particles from diffusing out.  

The low surface brightness X-ray diffuse emissions in front 
of the bow shock and outside of
the eastern edge of the PWN are positionally 
consistent with the bipolar H$\alpha$ structure (Figure 2 of Moon et
al. 2004). We propose that this is the consequence of the non-uniform
distribution of the SNR ejecta. In the directions of the two 
low surface brightness X-ray protrusions, the ejecta are of so low 
density that they can not effectively confine the wind particles and 
so the wind particles extend further out. Finally the wind particles 
are stopped by the ISM that has low metallicity, and the interaction 
between them ionizes the hydrogen in the ISM to be H$\alpha$ emitting.

\end{itemize}

\subsection{The origin of the apparent hard edge of the X-ray plateau}
Now we discuss the origin of the possible spectral hardening at the
plateau edge. Although, generally, the apparent harder spectrum  
can be the result of either a higher $N_{\rm H}$ or an
intrinsic hardening of the emitting spectrum,  
the first possibility can be ruled out for the hard edge of the 
X-ray plateau in this PWN. 
As shown in Table 2, with the photon index  ($\Gamma$) fixed, 
$N_{\rm H}$ to the plateau edge needs to be 
$4\pm3\times10^{20}$ $\rm cm^{-2}$  higher than those to the
neighboring regions. From the radio and optical observations 
we know that the shell has a width of $\sim$0.03 pc and an inner
 radius of about 0.1 pc. If the higher $N\rm_H$ to the plateau edge 
(and so the radio shell) is due to the additional absorbing material in the
shell, the required number density of the shell will be
$\sim(1600\pm1200)\rm cm^{-3}$,  about 20 times higher than 
50-100 $\rm cm^{-3}$, the value measured in 
optical observations \citep{HK1989}. This implies
that the apparent X-ray spectrum hardening of the edge of the plateau is not 
a consequence of the higher absorption but 
an intrinsic harder emitting spectrum. 

In turn, the intrinsic harder emitting spectrum may be  
due to either the re-acceleration of the pulsar wind particles at the edge or 
the contribution from a second emission component over there. 
For both cases, a shock wave is needed, and   
a shock wave at the interface between the PWN and SNR ejecta
does have already been proposed by \cite{HK1989} to explain 
the filamentary optical emission from the core of CTB 80.  This 
shock is in the SNR ejecta rather than in the PWN because of the high 
sound velocity ($c/\sqrt3$) in the PWN, and thus the shock can not 
re-accelerate the pulsar wind particles. However, it is not unlikely that 
this shock accelerates new relativistic particles and these particles
radiate synchrotron X-ray emission at the shock front, like those observed
from several SNRs \citep[e.g.,][]{Koyama1995}. This shock generalized
X-ray emission might have a flat spectrum, which makes the overall
spectrum of the X-ray emission from this region harder than those
from neighboring regions. 

\subsection{Constraints on the NS cooling models}
The cooling of NS (age $< 10^5$ yr) is realized mainly by
neutrino emission 
from the entire stellar body. The standard cooling model only 
includes  neutrino emission 
via the modified URCA process and the nonstandard models involves pion
(kaon) condensates, strong magnetic field, or neutron superfluidity
\citep[e.g.,][]{Page1998, Slane2002, Yakovlev2002, Yak2004}.

With the {\sl Chandra} observation we estimated
the effective blackbody temperature of the entire surface of PSR B1951+32
and compared it with the cooling models.
The majority of equations of state yield an effective NS radius larger
than 12 km for any range of masses  (Haensel 2001). In order to estimate
the blackbody temperature of PSR B1951+32, we fitted its X-ray spectrum with
a PL+BB model by fixing the emitting radius as 12 km, and obtained a 
blackbody temperature of $7.4\times10^5$ K with 3$\sigma$ upper limit of 
7.8$\times 10^5$ K. We plot in 
Figures 8 and 9 this temperature upper limit and the various
theoretical NS surface cooling curves. In the 
figures the pulsar's dynamic age (64$\pm$18 kyr) 
has been used, but the characteristic age of 107 kyr is not much larger than
the dynamic age \citep{Mig2002, Fruchter1988} and will not change the 
results much. 
Similar to \cite{Slane2002} and \cite{Halpern2004}, the
effective blackbody temperature falls 
considerably below the predictions of the standard cooling models,
suggesting 
the presence of some exotic cooling contribution (such as pion, 
kaon condensates or strong magnetic field effects) in the 
interior \citep{Slane2002, Yakovlev2002}. The influence of the 
mass to the cooling of the NS has been discussed by \cite{Yakovlev2002}.
 The larger mass NS tends to cool faster because the modified or direct 
URCA process are less suppressed by strongly proton superfluidity.
As seen from Figure 9,  our result may also indicate that the mass of
PSR B1951+32 is higher than 1.42M$_\sun$. 

In $\S$ 3.2 we pointed out that there is a 0.13 keV blackbody 
component in the X-ray spectrum of PSR B1951+32. However, this component
is not from the entire NS surface giving the small 
equivalent emitting
radius. In the above paragraph, this component had not been
excluded in deriving the temperature of the entire
surface of PSR B1951+32, and thus this surface temperature 
upper limit is modest. If the emission from the surface of a NS is not
a blackbody, then the modification by the presence of a whatever
atmosphere component other than H will not make the
temperature of the NS surface higher than the estimated effective temperature
upper limit \citep{Slane2002}. In short, the above upper limit of the
surface temperature of PSR B1951+32 and its constraints on the NS cooling
models are safe.

\section{Conclusions}       
We studied the morphology and spectra of PSR B1951+32 and its PWN.
The overall morphology of the PWN is consistent very well with 
that of a ram pressure confined PWN. The X-ray 
map shows 
a bright plateau which is within the radio and optical shells positionally,
and thus represents the main body of a PWN confined in
the SNR.   
We found that the spectrum of the  edge of the X-ray plateau is probably 
harder than those 
of the neighboring regions, which are possibly due to the shock
between the PWN 
and the SN ejecta. We detected thermal X-ray emission
from hot spots on the NS, and the temperature of the entire surface 
of PSR B1951+32 is shown to be much lower than that 
predicted by the standard NS cooling model.  
 
\acknowledgments
We thank the anonymous referee for his/her invaluable comments and suggestions
that led to the significant improvement of the paper. 
We also thank Profs. Mei Wu and Yong Chen of IHEP and Prof. Yang Chen of 
Nanjing University for discussions and suggestions.  This work is 
supported by the Special Funds for Major State Basic Research Projects and
the National Natural Science Foundation of China.

\begin{table*}[!htp]
\caption{The spectral fitting parameters of PSR B1951+32 and its
PWN}
\begin{center}
\begin{tabular}{lcr}
\noalign{\smallskip}
\hline
\hline
\noalign{\smallskip}
Region & $\Gamma$/T(keV)&Flux\\
\hline
\noalign{\smallskip}
Point Source  & \\
~~Powerlaw+blackbody&\multicolumn{2}{c}{}\\
\hspace{1cm}powerlaw       & $1.63\pm0.05$&  35$\pm$3\\
\hspace{1cm}blackbody      & $0.13\pm0.02$& $3^{+14}_{-3}$\\
~~Powerlaw       &  $1.74\pm0.03$&35$\pm$2\\
Bow shock     &  $1.6^{+0.1}_{-0.2}$&2.1$^{+0.9}_{-0.4}$\\
Tail Emission &  $1.6\pm0.1$&3.6$\pm$0.7\\
Plateau  &  $1.69\pm0.04$&61$\pm$4\\
Southwestern Diffuse Emission& $1.77^{+0.09}_{-0.07}$&8.0$\pm$0.9\\
Eastern Diffuse Emission & $1.8\pm0.1$&5.4$\pm$0.7\\
Outer Diffuse  Emission&  $1.88^{+0.08}_{-0.07}$&13$\pm$1\\
Entire nebula &1.73$\pm$0.03&96$\pm$5\\
\noalign{\smallskip}
\hline
\hline
\noalign{\smallskip}
\end{tabular}
\tablecomments{The X-ray absorbing column density $N_{\rm H}$ is obtained by jointly
fitting the spectra of various diffuse components  as
$(3.0\pm0.1)\times$10$^{21}$cm$^{-2}$. We rebinned the spectra before 
fitting so that data in each rebinned channel
have signal to noise ratio $\ge$ 6. The overall $\chi^2$ is 401 for
544 dof for the diffuse components. 
The unabsorbed flux density ($10^{-13}\rm ergs\,cm^{-2}\,s^{-1}$) in 
0.2-10 keV is listed in the 3rd column.}
\end{center}
\end{table*}

\begin{table*}[!htp]
\caption{The spectral fitting results of the 5 quasi-annulus regions}
\begin{center}
\begin{tabular}{l@{\hspace{0.15cm}}c@{\hspace{0.15cm}}r|c@{\hspace{0.15cm}}r}
\noalign{\smallskip}
\hline
\hline
%\noalign{\smallskip}
 &\multicolumn{2}{c}{$N_{H}=3.0\times
10^{21}$cm$^{-2}$}&\multicolumn{1}{|c}{$\Gamma=1.70\pm0.05$}\\
\hline
Regions & $\Gamma$&Flux & $N_{\rm H}(10^{21}\;$cm$^{-2})$\\

\hline
%\noalign{\smallskip}
Center  & $1.76\pm0.06$ & 10$\pm$1&$2.8\pm0.2$ \\
Ring 1  & $1.66\pm0.04$ & 20$\pm$1&$3.2\pm0.2$ \\
Ring 2  & $1.65\pm0.05$ & 14$\pm$1&$3.2\pm0.2$ \\
Ring 3  & $1.66\pm0.10$ & 10$\pm$1&$3.1\pm0.2$ \\
Ring 4  & $1.81\pm0.08$ & 8$\pm$1&$2.6\pm0.3$ \\
Ring 123  & $1.64\pm0.02$ & 42$\pm$2&$3.3\pm0.2$ \\
%\noalign{\smallskip}
\hline
\hline
\noalign{\smallskip}
\end{tabular}
\tablecomments{Columns 2 and 3 are fitted photon indices ($\Gamma$) and 
unabsorbed fluxes ($10^{-13}\rm ergs\,cm^{-2}\,s^{-1}$) in 0.2-10 keV,
when $N_{\rm H}$ is fixed as $3.0\times10^{21}\rm cm^{-2}$, the
$\chi^2$ is 350 for 530 dof. Columns  
4 are fitted $N_{\rm H}$ (10$^{21}$cm$^{-2}$) and unabsorbed 
fluxes ($10^{-13}\rm ergs\,cm^{-2}\,s^{-1}$) by forcing all spectra (except Ring 123) to share the 
same $\Gamma$ (which is derived as 1.70$\pm$0.05), and 
the corresponding $\chi^2$ is 348 for 529 dof. The spectrum index of Ring 123 is obtained with $N_{\rm H}=3.0\times10^{21}\rm cm^{-2}$, the $\chi^2$ is 204 for 241 dof. The $N_{\rm H}$ of Ring 123 is obtained with $\Gamma$ = 1.70, and the  $\chi^2$ is 196 for 241 dof.}
\end{center}
\end{table*}

\hspace{20cm}

\begin{figure}
\includegraphics[scale=.90]{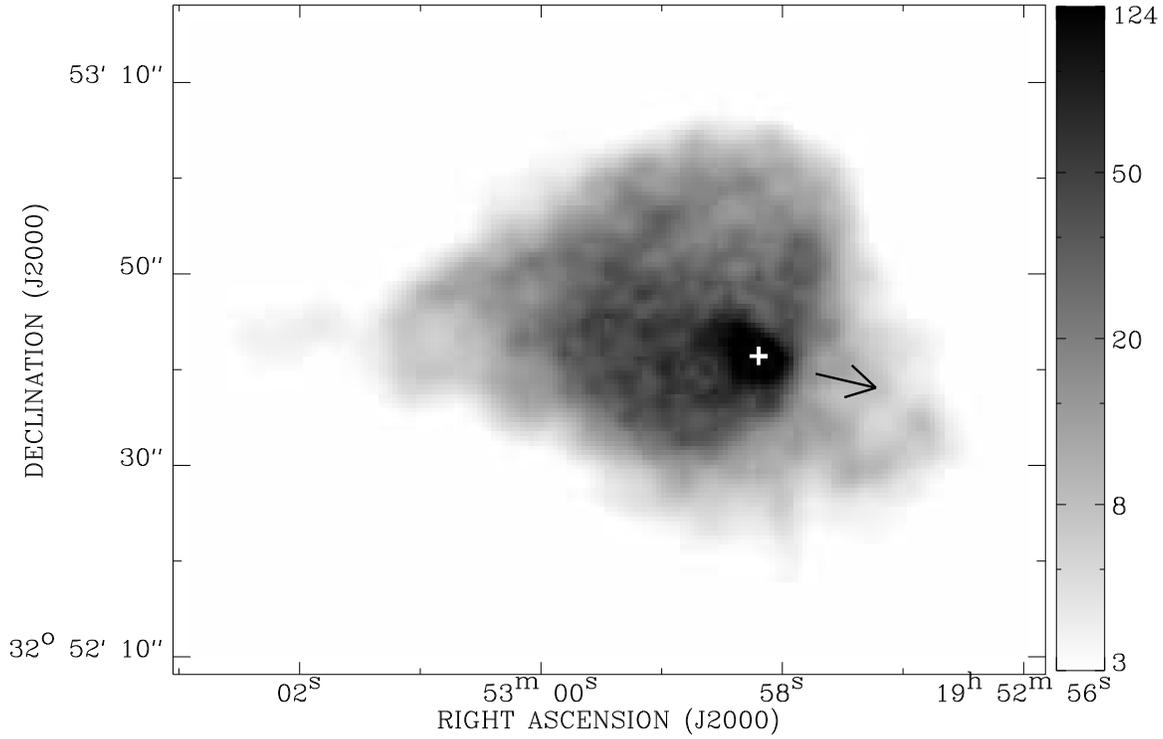}
\caption{The {\it Chandra} ACIS image of the PWN of PSR B1951+32,
adaptively smoothed with a Gaussian filter to achieve a signal to noise ratio $\ge$
12. The cross indicates the position of PSR B1951+32 and the arrow 
indicates the direction of the pulsar proper motion. The 
grey scale increase logarithmically from 3.3 to 124 counts per square arcsecond 
after smoothing.} 
\end{figure}

\begin{figure}
\includegraphics[scale=.90]{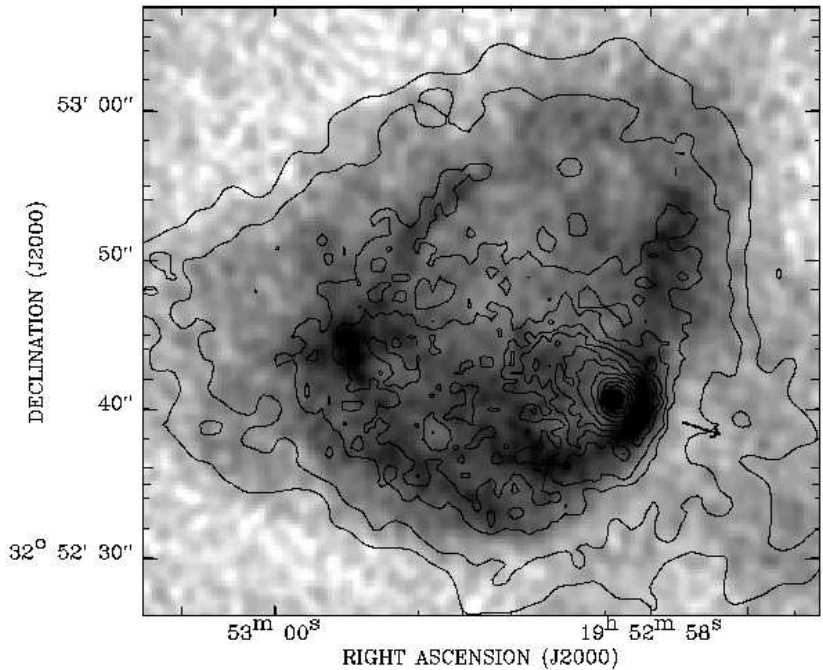}
\caption{The 0.3-8.0 keV X-ray contours (levels:  
12, 29, 41, 62, 83, 124, 206, 413, 826, 1652, 3304 counts per square arc second after
smoothing) of the PWN of PSR B1951+32 superposed upon the 20 cm radio image from
\cite{Mig2002}. The arrow indicates the direction of pulsar proper motion.}  
\end{figure}

\begin{figure}
\includegraphics[scale=.50,angle=-90]{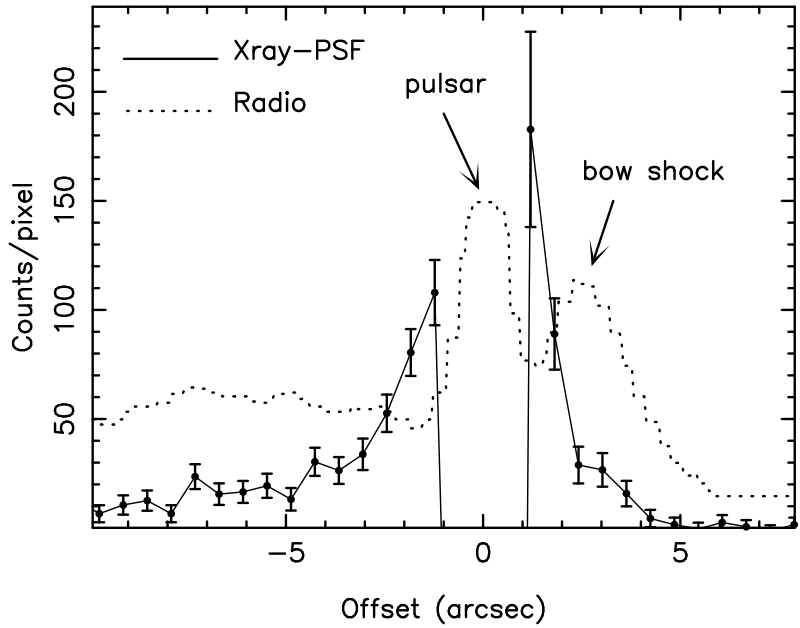}
\caption{The 0.3-8.0 keV X-ray intensity profile (solid
line, with the pulsar contribution subtracted, pixel size is
0$\farcs492\times0\farcs492$) and the radio profile (dashed line) 
along the proper motion direction of PSR B1951+32. The radio data are
taken from Figure 2 of Migliazzo et al. (2002) and 
plotted on an arbitrary linear scale. 
The x-axis represents the offset from the pulsar.} 
\end{figure}

\begin{figure}
\includegraphics[scale=.60]{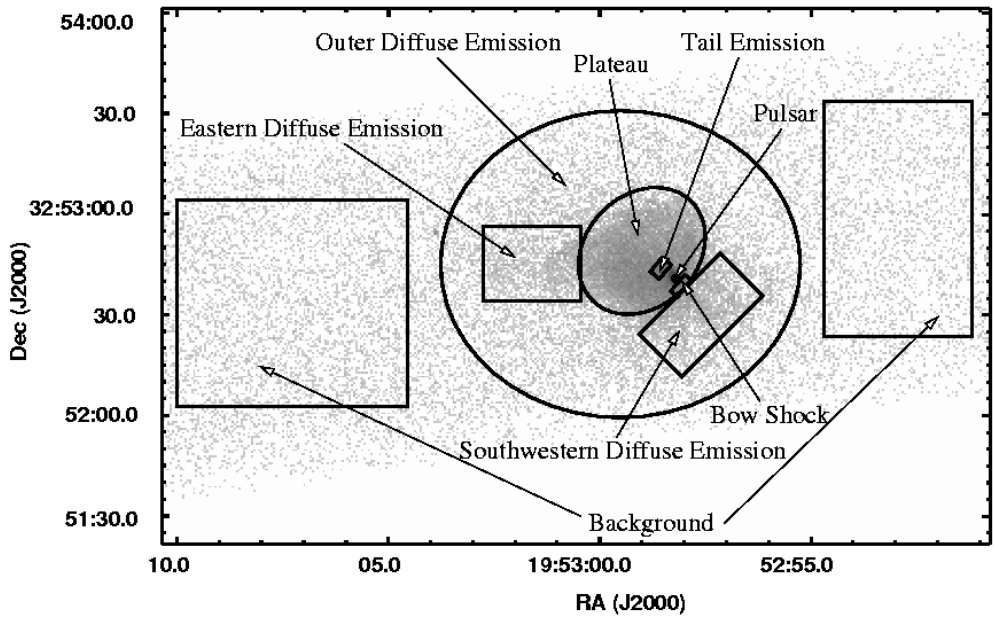}
\caption{The {\it Chandra} ACIS image of the PWN of PSR B1951+32. The
regions show where the spectra are extracted and the 
fitted parameters are listed in Table 1} 
\end{figure}

\begin{figure}
\includegraphics[scale=.60,angle=-90]{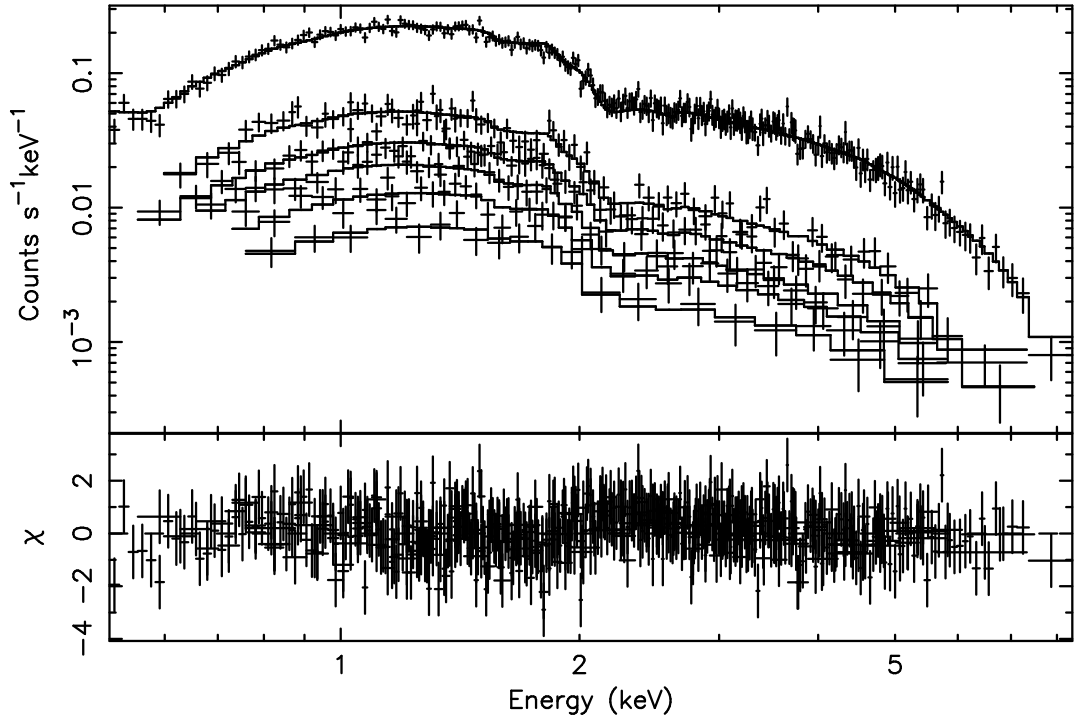}
\caption{Power law model fits to the spectra of different regions 
defined in Figure 4. From the top to the bottom, 
the spectra are of the plateau, outer diffuse emission, southwestern diffuse emission, eastern diffuse emission, tail emission, and bow shock regions.} 
\end{figure}
 \begin{figure}
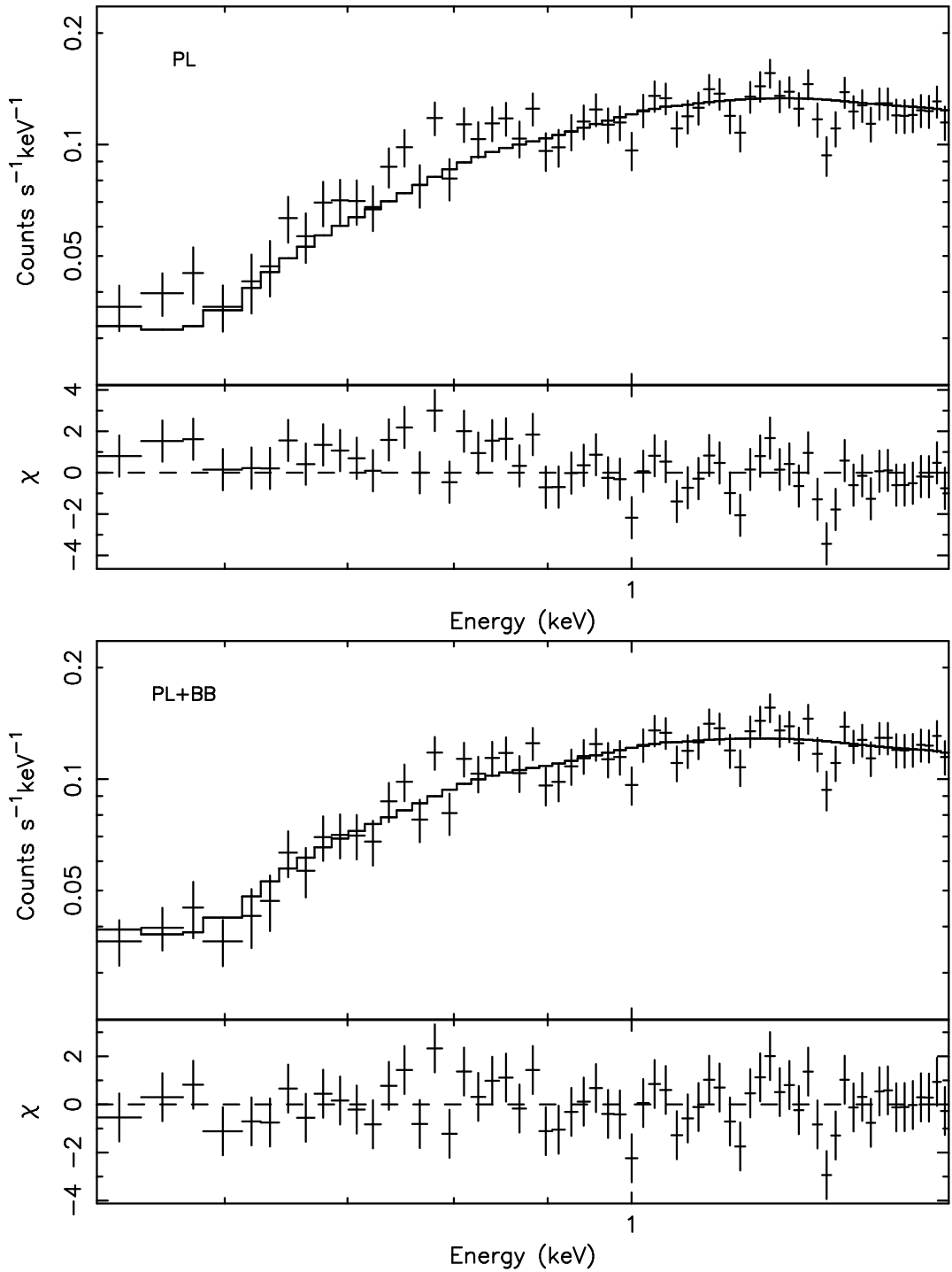

\includegraphics[scale=.60,angle=-90]{f6a.ps}
\includegraphics[scale=.60,angle=-90]{f6b.ps}
\caption{The spectral fitting to PSR B1951+32. The figures show only
data in 0.5-1.5 keV. The upper panel shows
the fit with a PL model and the lower panel with
a PL + BB model.} 
\end{figure}

\begin{figure}
\label{rings}
\includegraphics[scale=.60]{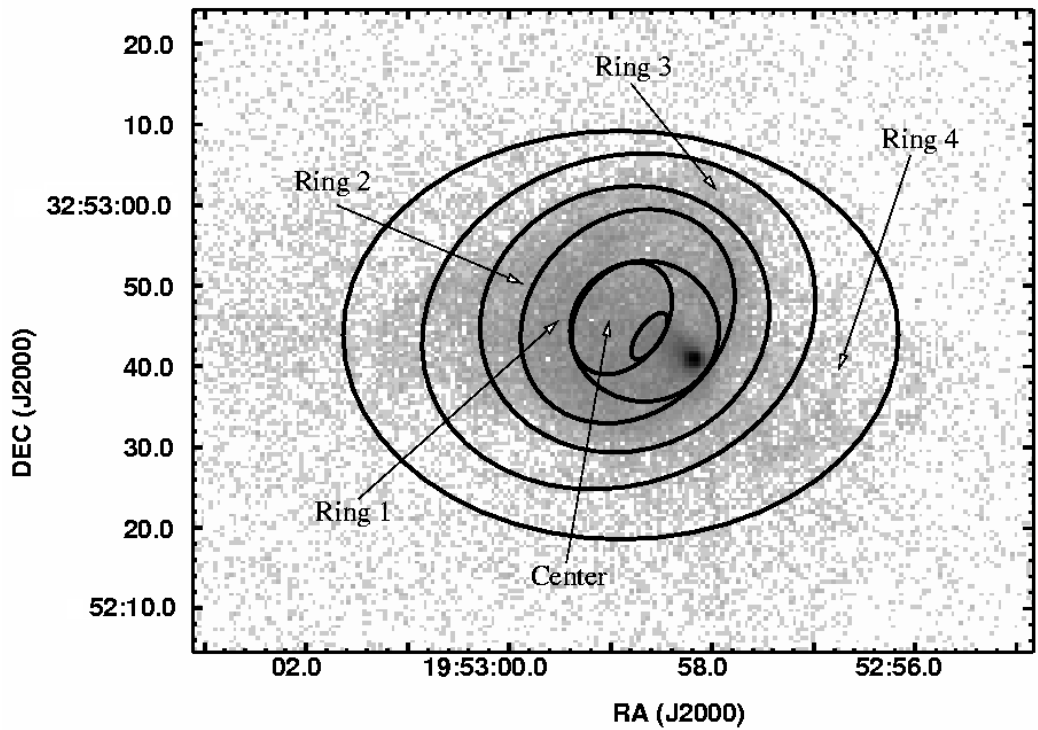}
\caption{The {\it Chandra} ACIS image of the PWN of PSR B1951+32. 
The quasi-annuli in this figure define regions from which the 
spectra listed in Table 2 are extracted.} 
\end{figure}

\begin{figure}
\includegraphics[scale=0.6]{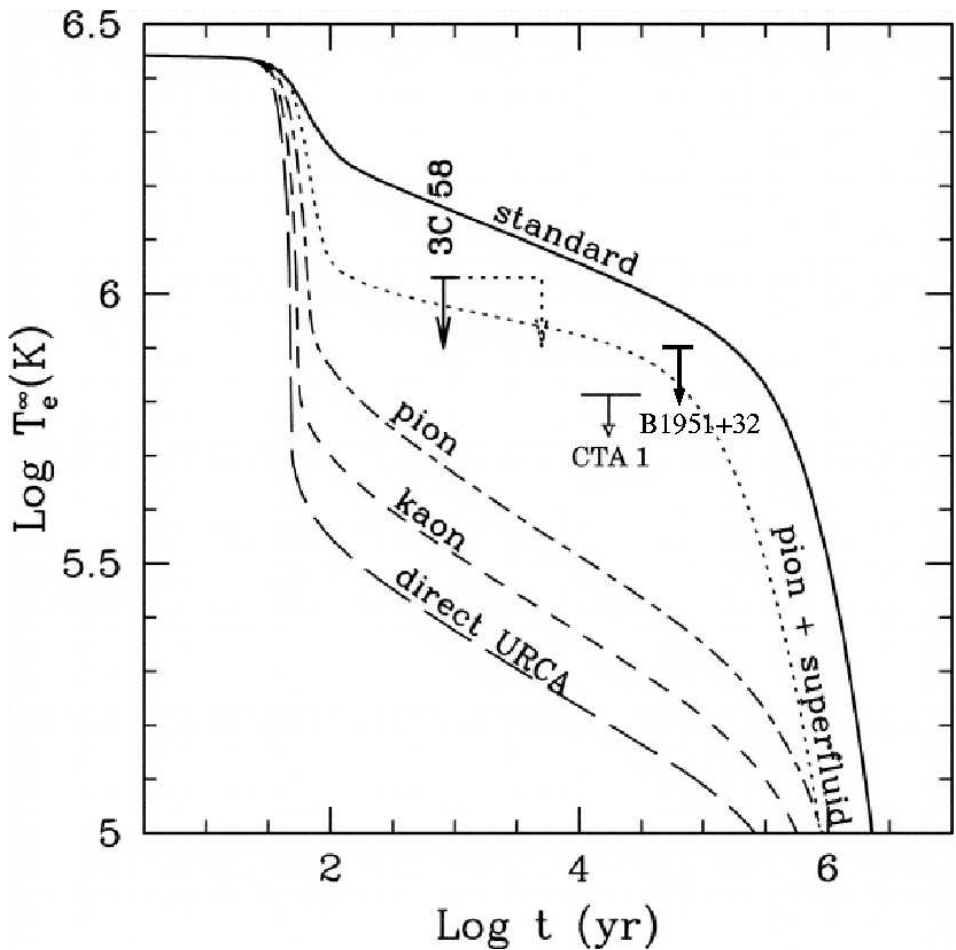}
\caption{The cooling models of a 1.4 $M_\odot$ neutron star
(Page 1998). PSR B1951+32 is plotted with the filled dark arrow. This
  figure is reproduced from Slane et al. (2002) and Halpern et
  al. (2004) with their permission. 
}
\end{figure}
\begin{figure}
\includegraphics[scale=0.60]{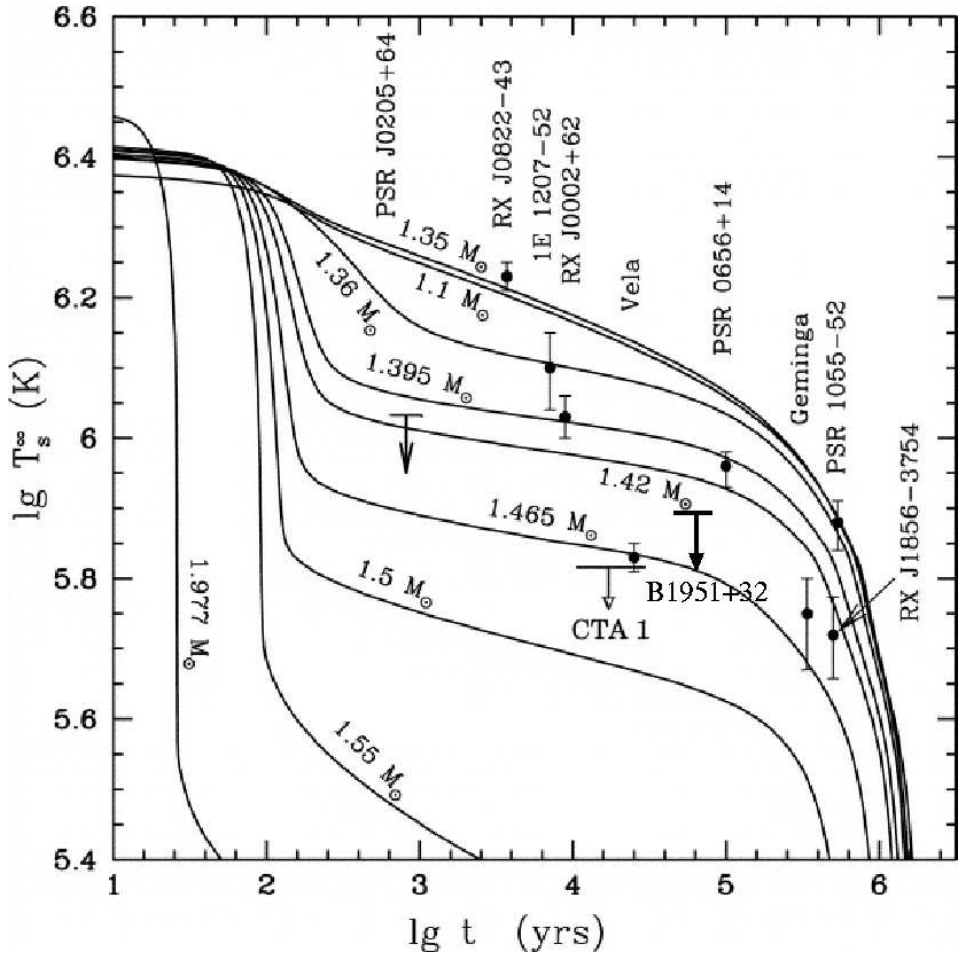}
\caption{The 
cooling models calculated assuming strong proton superfluidity and
weak neutron superfluidity (Yakovlev et al. 2002). PSR B1951+32 is
plotted with the filled dark arrow. This figure is reproduced from Yakovlev et
al. (2002) and Halpern et al. (2004) with their permission.
} 
\end{figure}


\begin{thebibliography}{}
\bibitem[Angelini et al.(1988)]{Angelini1988} Angelini, L., White, 
N.~E., Parmar, A.~N., Smith, A., \& Stevens, M.~A.\ 1988, \apjl, 330, L43 

\bibitem[Angerhofer et al.(1981)]{Angerhofer1981} 
Angerhofer, P.~E., Strom, R.~G., Velusamy, T., \& Kundu, M.~R.\ 1981, \aap, 
94, 313 

\bibitem[Becker et al.(1982)]{Becker1982} Becker, R.~H., Helfand, 
D.~J., \& Szymkowiak, A.~E.\ 1982, \apj, 255, 557 

\bibitem[Bucciantini(2002)]{Bucciantini2002} Bucciantini, N.\ 2002,
A\&A, 387, 1066

\bibitem[Caraveo et al.(2003)]{Caraveo2003} Caraveo, P.~A., 
Bignami, G.~F., DeLuca, A., Mereghetti, S., Pellizzoni, A., Mignani, R., 
Tur, A., \& Becker, W.\ 2003, Science, 301, 1345

\bibitem[Castelletti et al.(2003)]{Cas2003} Castelletti, G., 
Dubner, G., Golap, K., Goss, W.~M., Vel{\' a}zquez, P.~F., Holdaway, M., \& 
Rao, A.~P.\ 2003, \aj, 126, 2114 

\bibitem[Chatterjee \& Cordes(2002)]{Chatterjee2002} Chatterjee, S.~\& 
Cordes, J.~M.\ 2002, \apj, 575, 407

\bibitem[Frail et al.(1996)]{Frail1996} Frail, 
D.~A., Giacani, E.~B., Goss, W.~M., \& Dubner, G.\ 1996, \apjl, 464,
L165 

\bibitem[Fruchter et al.(1988)]{Fruchter1988} Fruchter, A.~S., 
Taylor, J.~H., Backer, D.~C., Clifton, T.~R., \& Foster, R.~S.\ 1988, \nat, 
331, 53 

\bibitem[Gaensler et al.(2002)]{Gaensler2002} Gaensler, B.~M., 
Arons, J., Kaspi, V.~M., Pivovaroff, M.~J., Kawai, N., \& Tamura, K.\ 2002, 
\apj, 569, 87

\bibitem[Gaensler et al.(2004)]{Gaensler2004} Gaensler, B.~M., van 
der Swaluw, E., Camilo, F., Kaspi, V.~M., Baganoff, F.~K., Yusef-Zadeh, F., 
\& Manchester, R.~N.\ 2004, \apj, 616, 383 

\bibitem[Gvaramadze(2004)]{Gvaramadze2004} Gvaramadze, V.~V.\ 2004, A\&A, 415, 1073

\bibitem[Haensel(2001)]{Haensel2001} Haensel, P.\ 2001, A\&A, 380, 186

\bibitem[Halpern et al.(2004)]{Halpern2004} Halpern, J.~P., 
Gotthelf, E.~V., Camilo, F., Helfand, D.~J., \& Ransom, S.~M.\ 2004, \apj, 
612, 398 

\bibitem[Hester \& Kulkarni(1988)]{HK1988} Hester, J.~J.~\& 
Kulkarni, S.~R.\ 1988, \apjl, 331, L121 

\bibitem[Hester \& Kulkarni(1989)]{HK1989} Hester, J.~J.~\& 
Kulkarni, S.~R.\ 1989, \apj, 340, 362 

\bibitem[Kaspi et al.(2001)]{Kaspi2001} Kaspi, V.~M., Gotthelf, 
E.~V., Gaensler, B.~M., \& Lyutikov, M.\ 2001, \apjl, 562, L163 

\bibitem[Kaspi et al.(2003)]{Kaspi2003} Kaspi, V.~M., Roberts, M.~S.~E., \& Harding, A.~K.\ 2004, astro-ph/0402136

\bibitem[Kennel \& Coroniti(1984)]{Kennel1984} Kennel, C.~F., \& 
Coroniti, F.~V.\ 1984, \apj, 283, 694 

\bibitem[Koo et al.(1993)]{Koo1993} Koo, B., Yun, M., 
Ho, P.~T.~P., \& Lee, Y.\ 1993, \apj, 417, 196 

\bibitem[Koyama et al.(1995)]{Koyama1995} Koyama, K., Petre, R., 
Gotthelf, E.~V., Hwang, U., Matsuura, M., Ozaki, M., \& Holt, S.~S.\ 1995, 
\nat, 378, 255 

\bibitem[Lu et al.(2002)]{Lu2002} Lu, F.~J., Wang, Q.~D., 
Aschenbach, B., Durouchoux, P., \& Song, L.~M.\ 2002, \apjl, 568, L49 

\bibitem[Lu et al.(2003)]{Lu2003} Lu, F.~J., Wang, Q.~D., \& Lang, C.~C.\ 2003, \aj,
126, 319

\bibitem[Mavromatakis et al.(2001)]{Mav2001} Mavromatakis, F., Ventura, J., 
Paleologou, E.~V., \& Papamastorakis, J.\ 2001, \aap, 371, 300 

\bibitem[Migliazzo et al.(2002)]{Mig2002} Migliazzo, J.~M., 
Gaensler, B.~M., Backer, D.~C., Stappers, B.~W., van der Swaluw, E., \& 
Strom, R.~G.\ 2002, \apjl, 567, L141 

\bibitem[Moon et al.(2004)]{Moon2004} Moon, D.-S., et al.\ 2004, 
\apjl, 610, L33 

\bibitem[Olbert et al.(2001)]{Olbert2001} Olbert, C.~M., 
Clearfield, C.~R., Williams, N.~E., Keohane, J.~W., \& Frail, D.~A.\ 2001, 
\apjl, 554, L205 

\bibitem[Page(1998)]{Page1998} Page, D. \ 1998, in The Many Faces of
neutron Stars, ed. R. Buccheri, J. van Paradijs, \& M. A. Alpar
(Dordrecht: Kluwer), 539

\bibitem[Petre et al.(2002)]{Petre2002} Petre, R., Kuntz, K.~D., 
\& Shelton, R.~L.\ 2002, \apj, 579, 404 

\bibitem[Pacholczyk(1970)]{Pacholczyk1970} Pacholczyk, A.~G.\ 1970, 
Series of Books in Astronomy and Astrophysics, San Francisco: Freeman, 1970

\bibitem[Rees \& Gunn(1974)]{Rees1974} Rees, M.~J.~\& Gunn, 
J.~E.\ 1974, \mnras, 167, 1 

\bibitem[Reynolds \& Chevalier(1984)]{Reynolds1984} Reynolds, S.~P.~ \& Chevalier, R.~A.~, 1984, \apj, 278, 630 

\bibitem[Safi-Harb et al.(1995)]{Safi_harb1995} 
Safi-Harb, S., Ogelman, H., \& Finley, J.~P.\ 1995, \apj, 439, 722 

\bibitem[Slane et al.(2002)]{Slane2002} Slane, P.~O., Helfand, 
D.~J., \& Murray, S.~S.\ 2002, \apjl, 571, L45 

\bibitem[Stappers et al.(2003)]{Stappers2003} Stappers, B.~W., 
Gaensler, B.~M., Kaspi, V.~M., van der Klis, M., \& Lewin, W.~H.~G.\ 2003, 
Science, 299, 1372 

\bibitem[Strom(1987)]{Strom1987} Strom, R.~G.\ 1987, \apjl, 319, 
L103 

\bibitem[van der Swaluw et al.(2003)]{Swaluw2003} van der Swaluw, 
E., Achterberg, A., Gallant, Y.~A., Downes, T.~P., \& Keppens, R.\ 2003, 
\aap, 397, 913 

\bibitem[van der Swaluw(2004)]{Swaluw2004} van der Swaluw, E.\ 2004, AdSpR,
33, 475

\bibitem[Wang et al.(1993)]{Wang1993} Wang, Q.~D., Li, 
Z., \& Begelman, M.~C.\ 1993, \nat, 364, 127 

\bibitem[Wang \& Gotthelf(1998)]{Wang1998} Wang, Q.~D.~\& 
Gotthelf, E.~V.\ 1998, \apj, 494, 623 

\bibitem[Wang et al.(2001)]{Wang2001} Wang, 
Q.~D., Gotthelf, E.~V., Chu, Y.-H., \& Dickel, J.~R.\ 2001, \apj, 559, 275 

\bibitem[Wang \& Seward(1984)]{Wang1984} Wang, Z.~R.~\& Seward, 
F.~D.\ 1984, \apj, 285, 607 

\bibitem[Weisskopf et al.(2000)]{Weisskopf2000} Weisskopf, M.~C., et 
al.\ 2000, \apjl, 536, L81 

\bibitem[Wilkin(1996)]{Wilkin1996} Wilkin, F.~P.\ 1996, \apjl, 
459, L31 

\bibitem[Yakovlev et al.(2002)]{Yakovlev2002} Yakovlev, D.~G., 
Kaminker, A.~D., Haensel, P., \& Gnedin, O.~Y.\ 2002, \aap, 389, L24 

\bibitem[Yakovlev \& Pethick(2004)]{Yak2004} Yakovlev, D.~G., 
\& Pethick, C.~J.\ 2004, \araa, 42, 169

\end{thebibliography}
\end{document}